\documentclass[usenatbib,psfig]{mn2e}

\usepackage{epsfig}

\begin{document}


\title[Age constraints for galactic bars]{Stellar population constraints on the ages of galactic bars}
\author[P. A. James \& S. M. Percival]{P. A. James\thanks{E-mail:
P.A.James@ljmu.ac.uk } \& S. M. Percival\\
Astrophysics Research Institute, Liverpool John Moores University, IC2, Liverpool Science Park, 146 Brownlow Hill, Liverpool L3 5RF, UK\\
}

\date{Accepted . Received ; in original form }

\pagerange{\pageref{firstpage}--\pageref{lastpage}} \pubyear{2015}

\maketitle

\label{firstpage}

\begin{abstract}
We present a study of the stellar populations within the central regions of four nearby
barred galaxies, and use a novel technique to constrain the duration of bar activity.
We focus on the star formation `desert', a region within each of
these galaxies where star formation appears to have been suppressed by
the bar.  New H$\beta$ spectroscopic data are presented, and used to produce spectroscopic line indices which are compared with theoretical predictions from population synthesis models for simple stellar populations and temporally truncated star formation histories.  This analysis shows that the dearth of star formation activity in these regions appears to have been 
 continuing for at least 1 Gyr, with timescales of several Gyr indicated for two of the galaxies. This favours models in which strong bars can be long-lived features of galaxies, but our results also indicate a significant diversity in stellar population ages, and hence in the implied histories of bar activity in these four galaxies.
\end{abstract}


\section{Introduction}
\label{sec:intro}

Bars are a common feature of spiral galaxies, with
near-IR imaging revealing that the majority have some bar structure in
their central regions, while strong bars are seen in about one-third
of spirals \citep{knap00, mari07}.  Much
theoretical and observational effort has recently been devoted to the
possible role of bars in driving morphological change in their host
galaxies.  In particular, tidal torques from bars are thought to
lead to central concentration of disk gas, with the resulting star
formation causing the build-up of `pseudobulges' \citep[][and references therein]{korm04}.  However, the strongest evidence for
this process is from theoretical simulations, and the
direct observational evidence is not yet compelling.  A central issue
is the lifetime and duty cycle of bars: is a barred galaxy always a
barred galaxy, and if not, how long does the bar phase endure?
Theoretical results are not clear.

One aspect of this question is the long-term stability of bars in the presence of a central mass concentration such as a super-massive black hole.  The most comprehensive study into this question, by \citet{shen04} found that bars are robust to the effects of plausible central masses (up to several per cent of the disk mass), a result that was confirmed by the models of \citet{deba06}.  Some other studies contradict this conclusion, with models that indicate a possible role for central mass concentrations in dissolving bars rapidly \citep{bour02,bour05,hozu05,hozu12}.
 
Simulations by
\citet{bere07} and \citet{curi08} show that bars can be stable for at
least 5 and 10~Gyr respectively, while \citet{comb90} find characteristic bar lifetimes of a 
Hubble time or more. Other studies \citep[e.g.][]{lang14} suggest that bars may be induced by external phenomena such as galaxy -- galaxy tidal interactions, which would predict a broad spread of bar ages.

Observationally,
\citet{pere09} and  \citet{pere11} have attempted to address this question using
optical spectroscopy to determine ages and metallicities along bars in 20 
nearby early-type barred galaxies, finding a wide range in luminosity-weighted ages of
the stellar populations.  However, the age of the stellar
population is not simply related to the age of the bar itself, and indeed 
\citet{wozn07} presents models predicting significant age differences within 
a given bar due to trapping of young stars at orbital resonances. Thus,
 we suggest here a complementary technique, based on the influence of
the bar on the star formation (SF) activity in its surrounding environment,
rather than within the bar itself.

This paper builds on the results of a recent H$\alpha$ spectroscopic study of 15 nearby barred galaxies \citep[][henceforth JP15]{jame15}. These galaxies had previously been identified \citep{jame09} as having a specific pattern of emission in continuum-subtracted narrow-band H$\alpha$ imaging, with a pronounced dip in this emission in the radial range swept out by the bar.  This region was termed the `star formation desert' (SFD), a terminology we follow in the present paper. This behaviour is clearly apparent in the H$\alpha$ images presented for the 4 galaxies in the present study, in Figs.~1--4, all showing regions devoid of H$\alpha$ emission over a radial range that corresponds closely with the bar extent. 
This characteristic SF pattern is confirmed by other indicators, e.g. three of the galaxies studied here (all except for NGC~2712) have GALEX ultra-violet light profiles presented by \citet{gild07}, all of which show distinctive peaks or plateaus of red far-UV -- near-UV colours in the radial range corresponding to the SFD.  Further confirmation comes from \citet{hako15} who demonstrate a deficit of core-collapse supernovae in barred galaxies over the radial ranges corresponding to SFDs, again confirming the suppression of SF in these regions.

Note that, in adopting a functional definition of the regions studied in terms of star-formation properties, we deliberately side-step the thorny terminological and classification issues of whether we are studying bulges, outer bulges, pseudo-bulges, disks, lenses or `barlenses'.

The deep long-slit optical spectroscopy presented in JP15 shows that most of these SFD regions do in fact have diffuse H$\alpha$ line emission, at a very low level that is not picked up by narrow-band imaging. However, such emission was found always to have [NII]/H$\alpha$ line ratios very different from those found from star-formation regions; this and the extended diffuse morphology led us to conclude that the emission is not powered by SF, but more probably by post-Asymptotic Giant Branch stars or shocks from the bar. Such emission was also found, using rather different methods, in galaxies studied by \citet{sing13}. This suppression of SF over intermediate radial ranges was anticipated by \citet{reyn98}, who proposed bar-induced shocks as an efficient method for producing such suppression effects.

In this paper, we attempt to constrain the longevity of bar activity by an analysis of the SFD regions of 4 of the galaxies studied in JP15.  We present additional H$\beta$ long-slit spectroscopic observations for these galaxies, and compare measured absorption line strengths with predictions from population synthesis models to constrain the time since the cessation of SF activity.  One important aspect of this analysis is a robust and self-consistent treatment of the effects of line emission on observed absorption-line strengths, given our finding of near-ubiquitous but low-level line emission in these regions.

The structure of this paper is as follows. Section~\ref{sec:gal_obs} contains a description of the observed galaxies and their main properties, and lists the new spectroscopic observations used in this analysis.  In Section~\ref{sec:dat_red} we describe the data reduction that was performed to extract absorption line strengths from long-slit spectroscopy.  The data analysis presented in Section~\ref{sec:analysis} includes a description of the iterative procedure adopted to correct line indices for the effects of line emission, and presents the theoretical population synthesis models with which the corrected indices are compared. Section~\ref{sec:discussion} contains the discussion of the main results in terms of stellar population ages and their consequences for our understanding of bar activity.  The main conclusions are summarised in Section~\ref{sec:conc_fut_work}.

\section{Galaxy sample and observations}
\label{sec:gal_obs}

\begin{table*}
 \begin{minipage}{140mm}
  \caption{Properties of the observed galaxies.}
  \begin{tabular}{llrccrr}
  \hline
 Name      &  Classn.   &   Velocity    & Distance &  Inclination  &   Galaxy PA &  Bar PA  \\
           &            & ~~km~s$^{-1}$  &   Mpc    & $^{\circ}$    & $^{\circ}~~~$ & $^{\circ}$~~~ \\
   \hline            
NGC~2543   &  SB(s)b    &  2471      & 38.6 &    55       &    45~~~   &   90~~~  \\
NGC~2712   &  SB(r)b    &  1815      & 31.2 &    57       &   178~~~   &   28~~~  \\
NGC~3185   &  (R)SB(r)a &  1217      & 23.4 &    47       &   130~~~   &  113~~~  \\
NGC~3351   &  SB(r)b    &   778      & 10.2 &    47       &    13~~~   &  110~~~  \\
\hline
\end{tabular}
\label{tab:gals_props}
\end{minipage} 
\end{table*}

\begin{figure}
\includegraphics[width=75mm,angle=0]{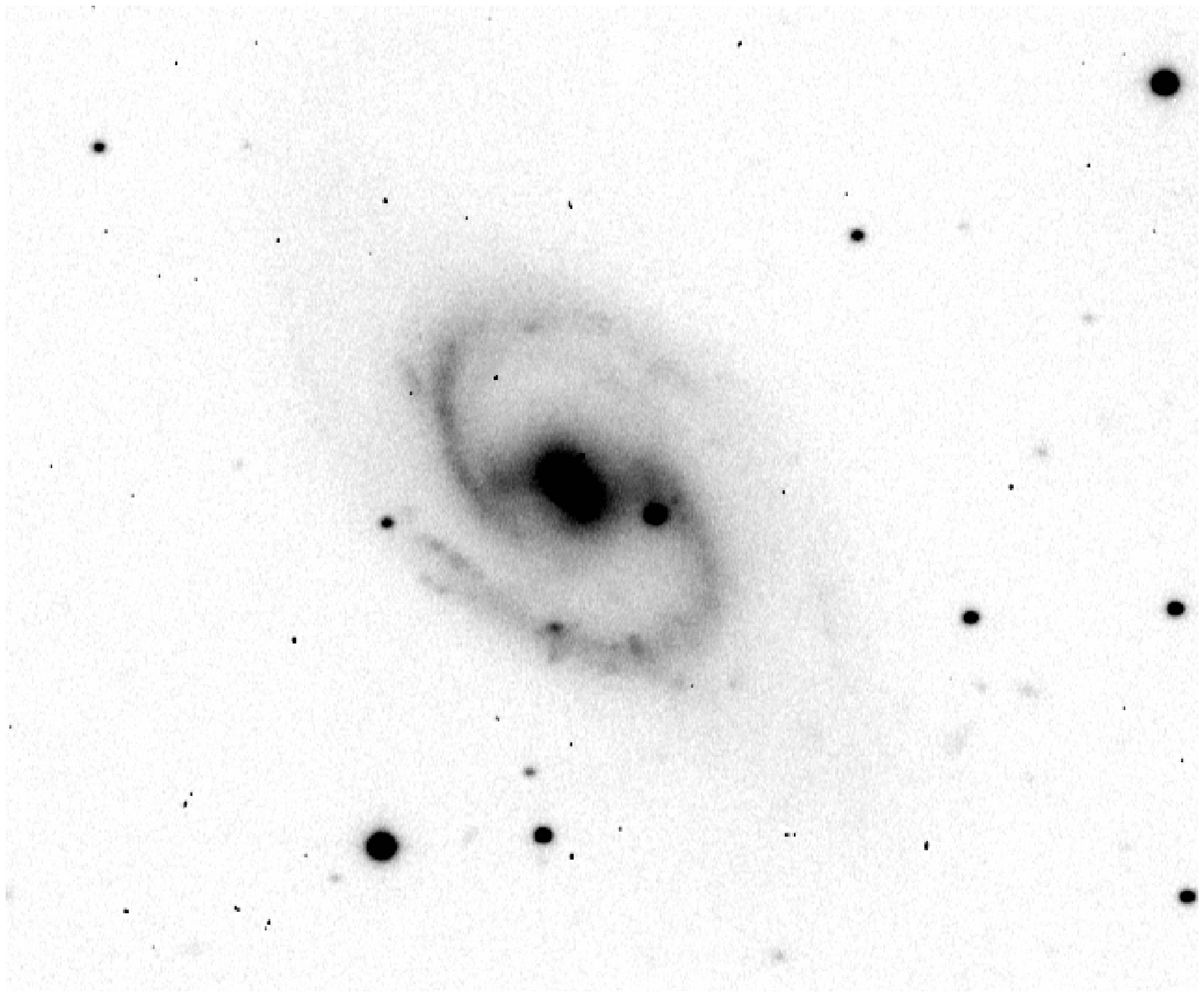}
\includegraphics[width=75mm,angle=0]{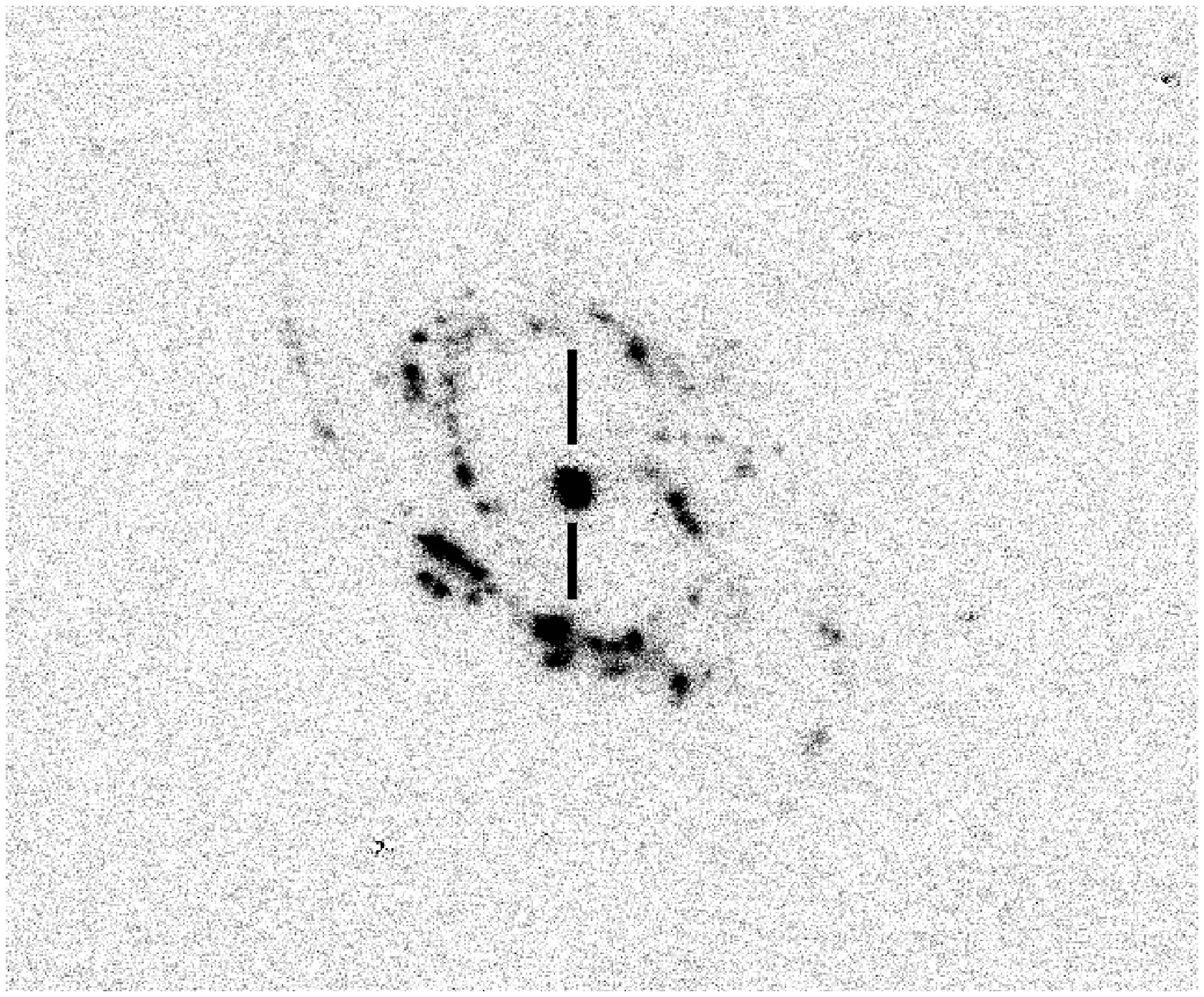}
\caption{NGC~2543, showing $R$-band (upper frame) and continuum-subtracted H$\alpha$ (lower frame) images. North is up and east to the left, and the image covers an area of 200~arcsec by 172~arcsec.}
\label{fig:U4273}
\end{figure}

\begin{figure}
\includegraphics[width=75mm,angle=0]{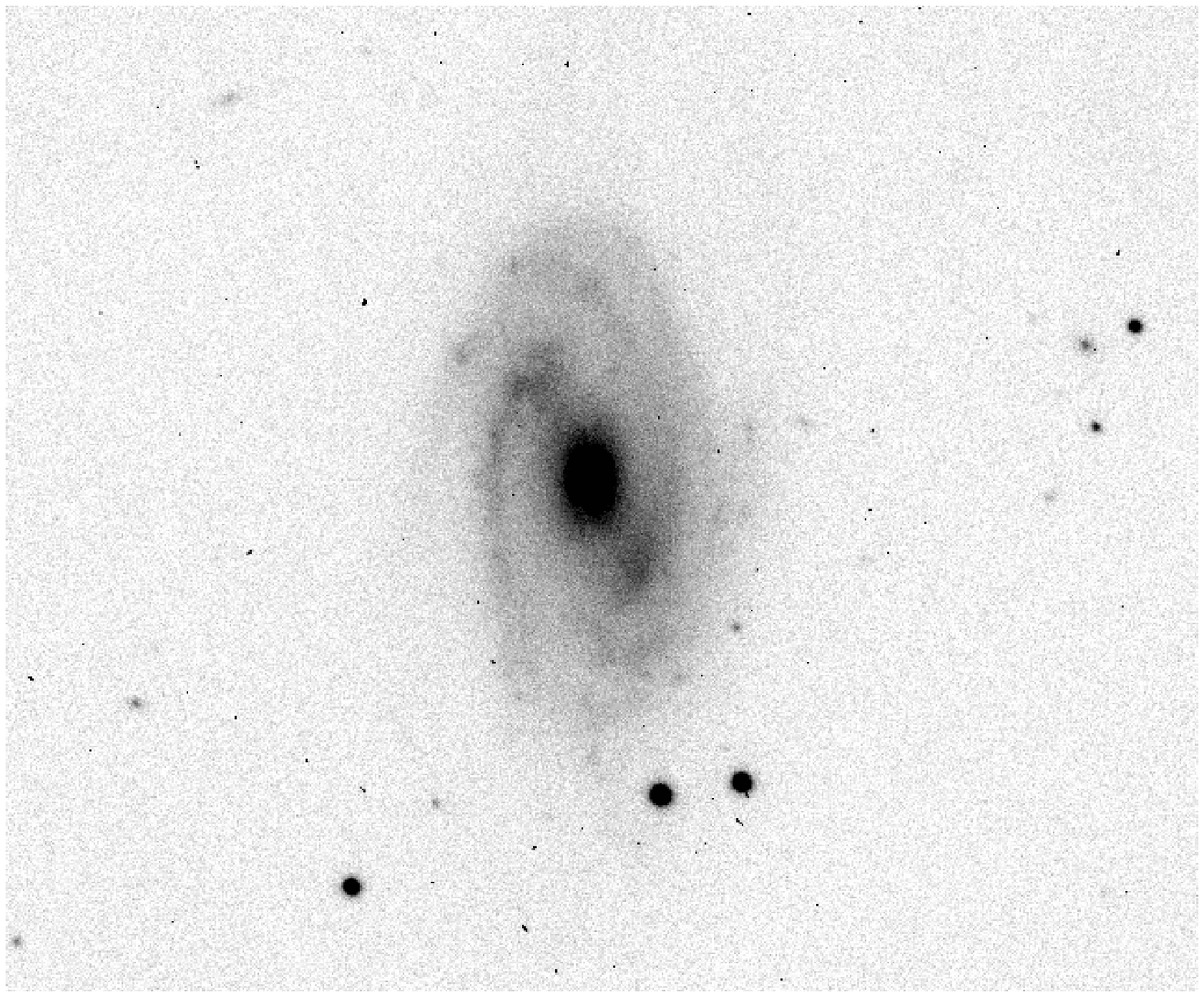}
\includegraphics[width=75mm,angle=0]{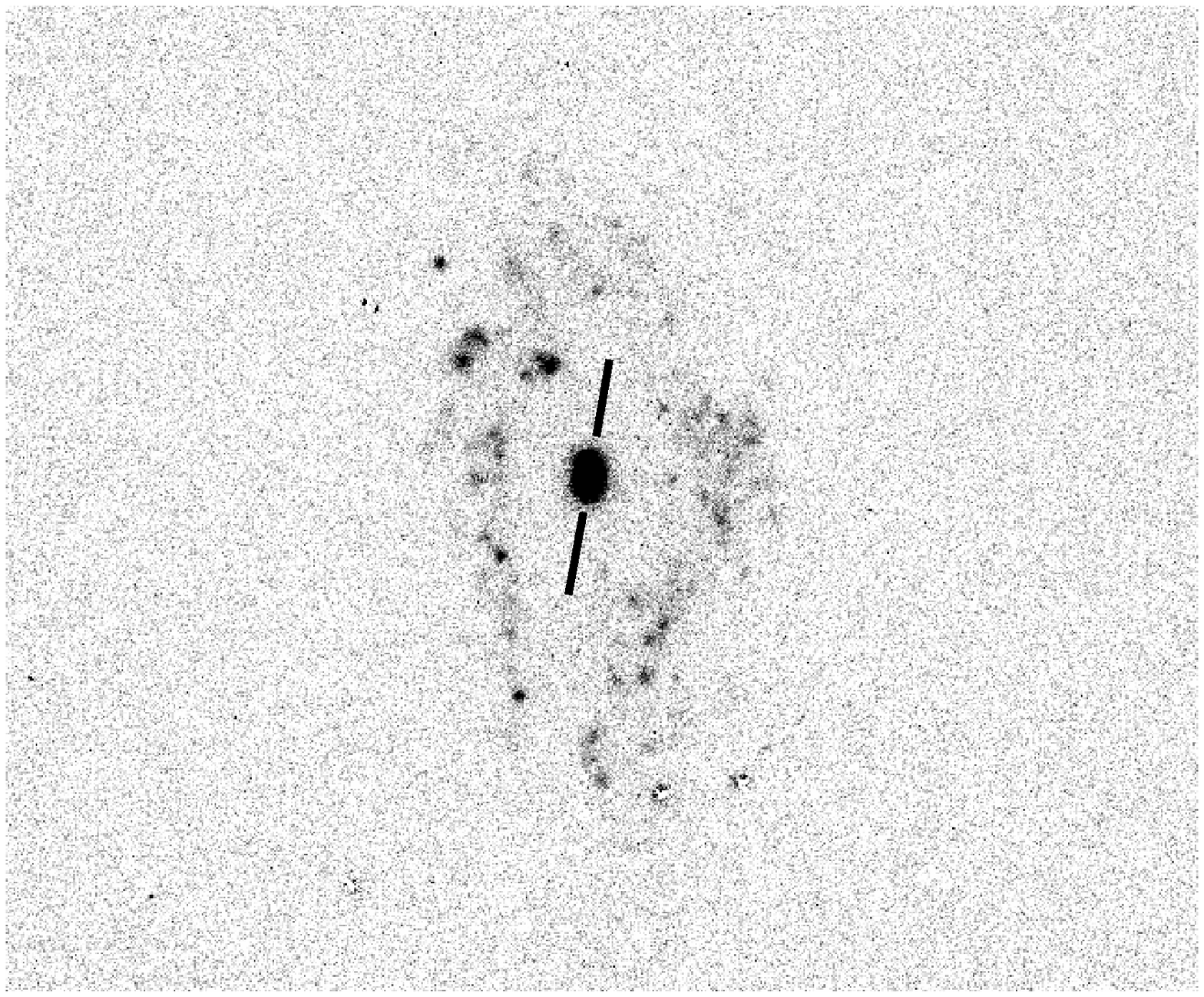}
\caption{NGC~2712, showing $R$-band (upper frame) and continuum-subtracted H$\alpha$ (lower frame) images. North is up and east to the left, and the image covers an area of 200~arcsec by 172~arcsec.}
\label{fig:U4708}
\end{figure}

\begin{figure}
\includegraphics[width=75mm,angle=180]{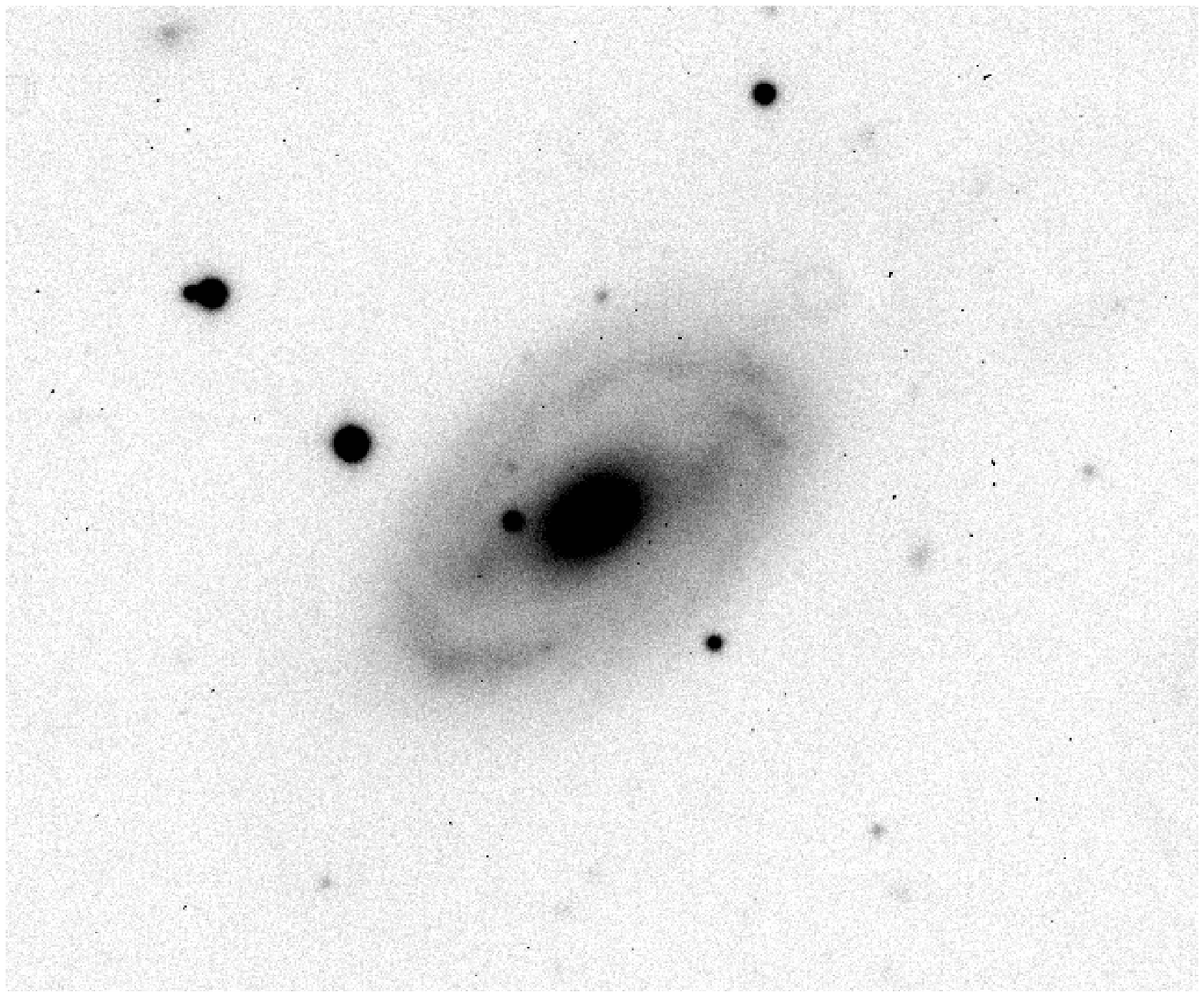}
\includegraphics[width=75mm,angle=180]{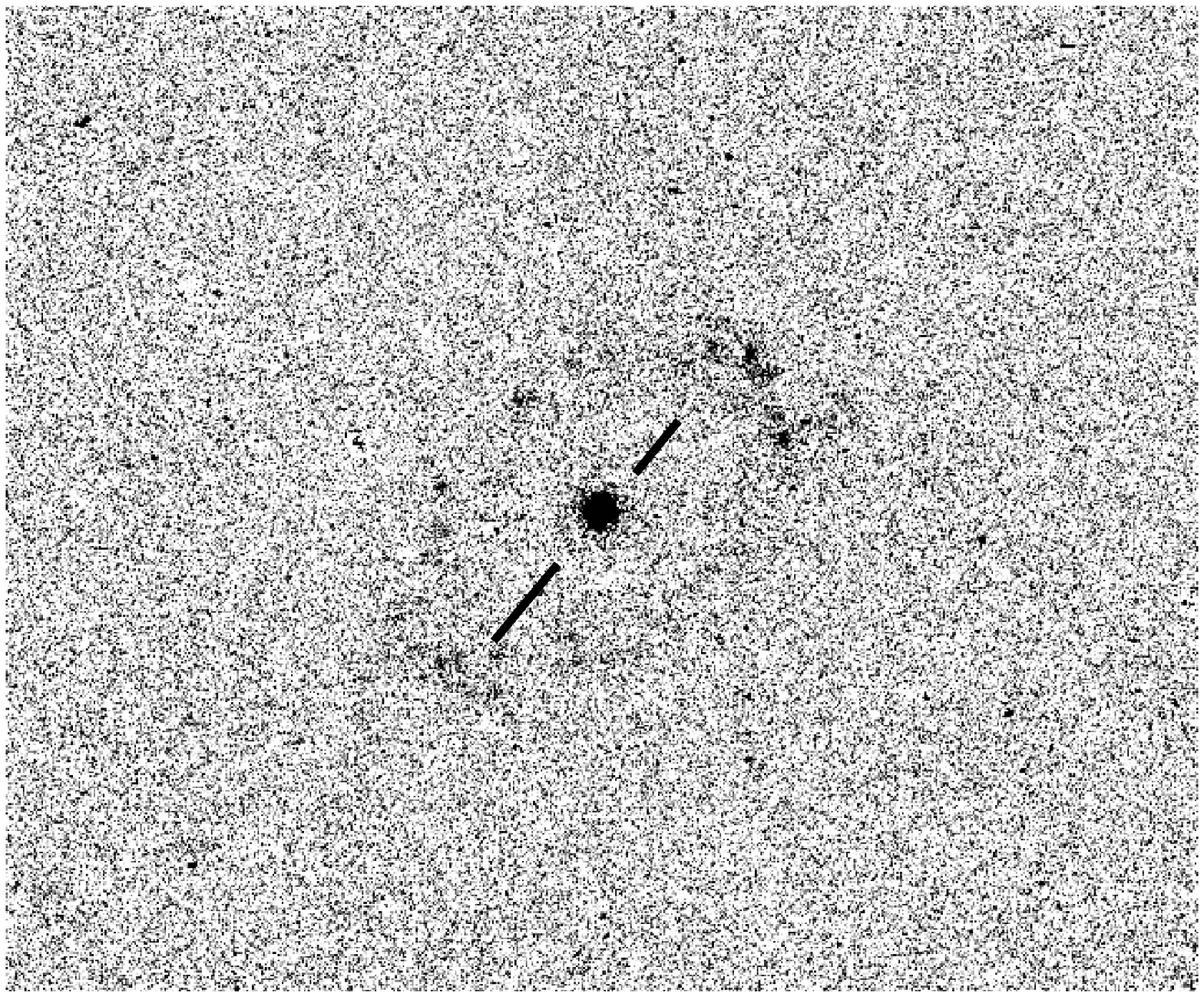}
\caption{NGC~3185, showing $R$-band (upper frame) and continuum-subtracted H$\alpha$ (lower frame) images. North is up and east to the left, and the image covers an area of 200~arcsec by 172~arcsec.}
\label{fig:N3185}
\end{figure}

\begin{figure}
\includegraphics[width=75mm,angle=0]{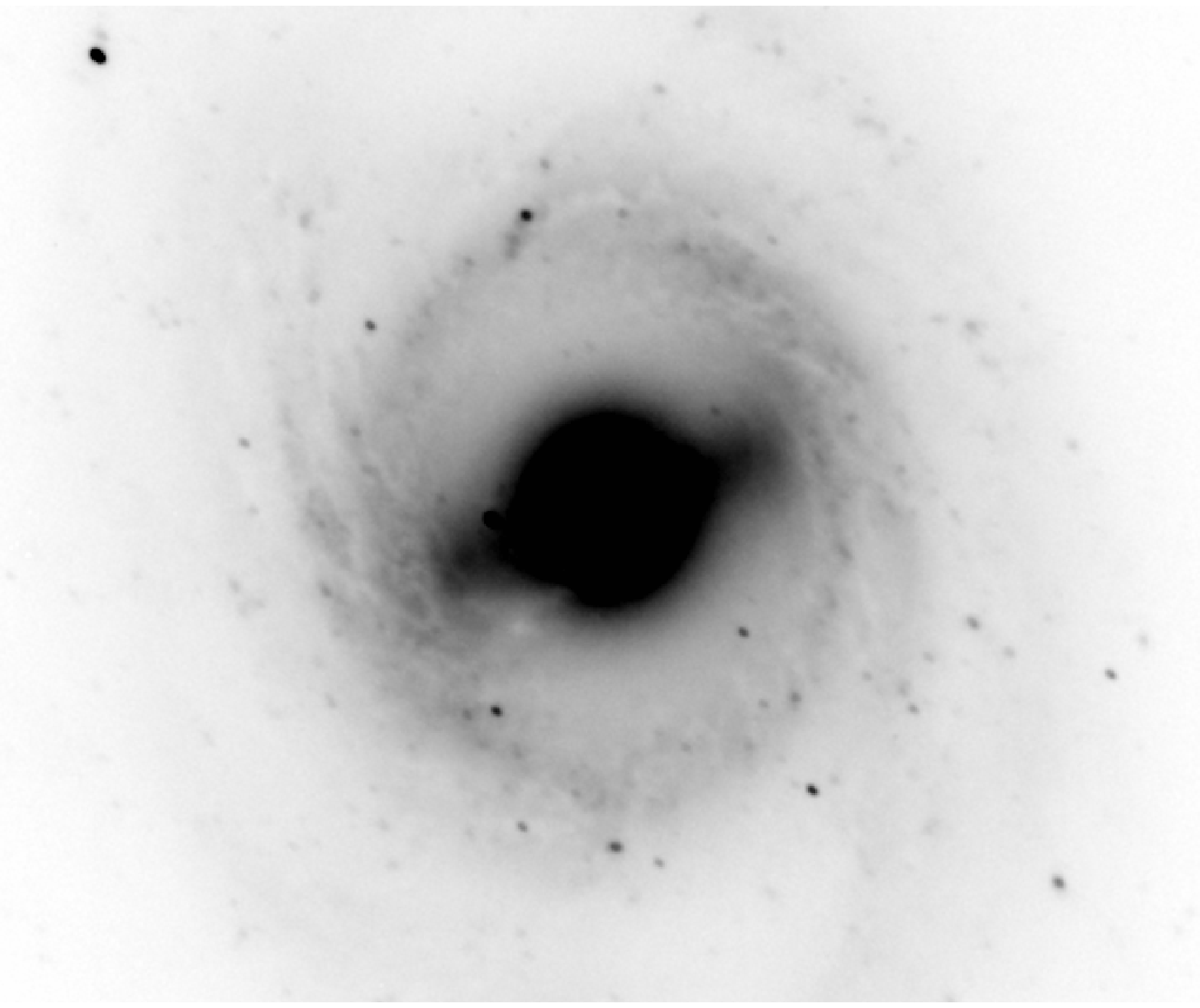}
\includegraphics[width=75mm,angle=0]{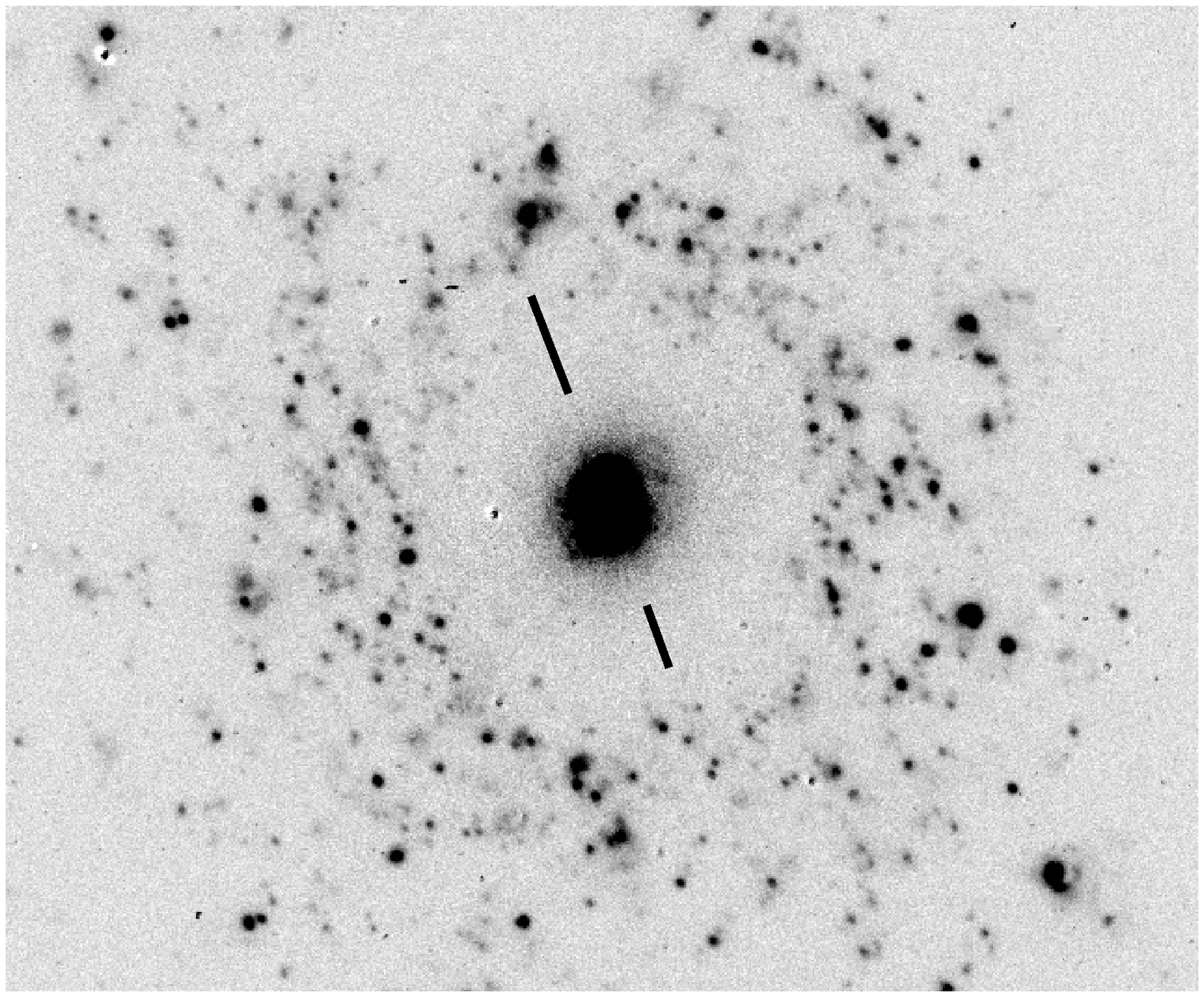}
\caption{NGC~3351, showing $R$-band (upper frame) and continuum-subtracted H$\alpha$ (lower frame) images. North is up and east to the left, and the image covers an area of 266~arcsec by 214~arcsec.}
\label{fig:N3351}
\end{figure}

The galaxies studied here are 4 low-redshift, strongly-barred galaxies, all of which are viewed at moderate angles of inclination.  Three are of type SBb, and one is classified as SBa. The main properties of the galaxies are listed in Table~\ref{tab:gals_props}, including distances (column 4) which are derived from a Virgo-infall corrected recession velocity for the most distant galaxy (NGC~2543), and from velocity-independent distance indicators such as Cepheid variables or type Ia supernovae for the other three.  Columns 5--7 in Table~\ref{tab:gals_props} list the galaxy inclination to the line-of-sight, the position angle of the galaxy major axis, and the position angle of the bar, measured from our own $R$-band imaging, in that order.

Two sets of spectroscopic observations are used in the analysis of these galaxies.  One set was taken with the Isaac Newton Telescope (INT) in the red end of the optical spectrum, specifically targeting the region including the H$\alpha$ line and the nearby [NII] lines.  The reduction and analysis of these spectra were described fully in JP15 and so will not be discussed further here. Spectra were extracted for two SFD regions in each of the four galaxies discussed here.  All eight of these regions showed evidence for low-level diffuse emission in the H$\alpha$ 6563~\AA\ and [NII] 6584~\AA\ lines, and in every case the ratios of the emission strength in these two lines were inconsistent with SF as the excitation mechanism.  The [NII]/H$\alpha$ ratios all lay in the range 0.51 -- 1.77, after application of a conservative correction for underlying H$\alpha$ absorption, whereas similarly corrected spectra from SF regions in the full galaxy sample showed a mean ratio of 0.330$\pm$0.013. Thus the conclusion from JP15 is that these SFD regions have strongly, and probably completely, suppressed SF activity, but that they do have diffuse line emission from some other source, which must be taken into account when interpreting absorption-line strengths in the present analysis.

\begin{table*}
 \begin{minipage}{140mm}
  \caption{Details of the long-slit spectroscopic observations taken with the INT and WHT.
}
  \begin{tabular}{lrccrccccrc}
  \hline
 Name      & Bar PA      &  Date   & Int time     &  Slit PA     & AM    & R$_{\rm MAX}$ & Date     & Int time & Slit PA      & AM\\
           & $^{\circ}$~~ &  INT    &    s         & $^{\circ}$~~~ &       &  kpc         & WHT      &   s      & $^{\circ}$~~~ &   \\
   \hline            
NGC~2543   &   90~~ &  20140204  &   1$\times$1200   &   0~~ & 1.01       &  4.5  &  20101227  & 5$\times$1200 &   0~~ & 1.02-1.13 \\
NGC~2712   &   28~~ &  20140207  &   3$\times$1200   & 170~~ & 1.04-1.05  &  3.3  &  20101227  & 5$\times$1200 & 170~~ & 1.04-1.05 \\
NGC~3185   &  113~~ &  20140204  &   3$\times$1200   & 140~~ & 1.01       &  3.3  &  20101227  & 3$\times$1200 & 150~~ & 1.01 \\
NGC~3351   &  110~~ &  20140205  &   3$\times$1200   &  20~~ & 1.05-1.06  &  2.5  &  20101227  & 4$\times$1200 &  20~~ & 1.05-1.12\\
\hline
\end{tabular}
\label{tab:gals_obs}
\end{minipage} 
\end{table*}

Additional spectroscopy, presented here for the first time, was taken using the William Herschel Telescope (WHT) with the Intermediate Dispersion Spectrograph and Imaging System (ISIS), in long-slit spectroscopy mode, on the night of 27 December 2010. The total exposure times were 1.0 - 1.7 hours per galaxy, as listed in column 10 of Table~\ref{tab:gals_obs}. 
ISIS was used with the R600B grating and a 
1.5$^{\prime\prime}$ wide slit, giving a spectral resolution of 
$\sim$ 2.5~\AA\ at the wavelength of the H$\beta$ 4861~\AA\ line which was the primary goal of the WHT spectroscopy. This resolution is sufficient to determine the spectral indices we require, 
from directly-measured equivalent width (EW) values within the bandpasses defined 
by \citet{trag98}, as described in Section~\ref{sec:analysis}.

Columns 5 and 10 of Table~\ref{tab:gals_obs} give the position angle (PA) on the sky of the slit for the INT and WHT observations respectively. For three of the four galaxies, the PA used was identical for the two sets of observations, while for the other,
NGC~3185, the slit PA differed by 10 degrees. It should also be noted that the spectra were not observed at the parallactic angle, and so atmospheric dispersion will lead to some differential flux loss as a function of wavelength.  The majority of the analysis presented here is based on EW measurements which are entirely unaffected by atmospheric dispersion effects, but the emission line correction described in Section~\ref{sec:analysis} does rely on H$\alpha$ to H$\beta$ emission line ratios and will be somewhat affected. However, it is important to note that all observations were taken at low airmass, with a maximum value of 1.13 and most observations being below airmass 1.06, so these atmospheric dispersion effects will be very small. For all observations with both the INT and WHT, the slit was positioned to pass directly through the nuclear peak of continuum emission as seen on the acquisition TV of the telescope.

\section{Data reduction}
\label{sec:dat_red}

Data reduction followed standard long-slit CCD spectroscopy procedures of bias subtraction, flat fielding, wavelength and flux calibration, and will not be described in detail here.  Flat fielding made use of tungsten arc spectra, while wavelength calibration was provided by copper-neon$+$copper-argon emission-line lamp spectra which were always taken immediately before or after the spectrum which they calibrated, at the same telescope pointing angle, to avoid differential flexure effects. Spectrophotometric standard star observations from the same night were used to provide a flux calibration.

For each of the four galaxies, two spectra were extracted, one from each side of the nucleus, by co-adding spectral information from appropriate parts of the slit.  The selection of these regions was described in JP15; the H$\beta$ spectra from the WHT were extracted in exactly the same way, using identical slit regions (although, given the slightly different slit PA for NGC~3185, the regions of the galaxy observed for the H$\alpha$ and H$\beta$ spectra are slightly offset).  These regions were defined to lie completely within the SFD of each galaxy, i.e. the radial range swept out by the bar and the slit regions are indicated by the black bars shown on the continuum-subtracted H$\alpha$ images in Figs. 1--4.  In all cases, no emission is apparent along the chosen slit positions in these H$\alpha$ images.  The spectra from opposite sides of the nucleus in each galaxy could have been co-added to improve signal-to-noise.  This was not done, partly to enable the investigation of spectral variation within galaxies, and partly because of data reduction complications resulting from velocity differences between these regions caused by galaxy rotation.  Thus, the end result of the basic data analysis was 8 flux-calibrated, sky-subtracted spectra for the SFD regions of the 4 barred galaxies, covering the spectral region around the H$\beta$ stellar absorption feature, and matching the 8 H$\alpha$-region spectra already described in Section~\ref{sec:gal_obs}.

\section{Data analysis}
\label{sec:analysis}

\subsection{Calculation of absorption line indices}

\begin{figure}
\includegraphics[width=80mm,angle=0]{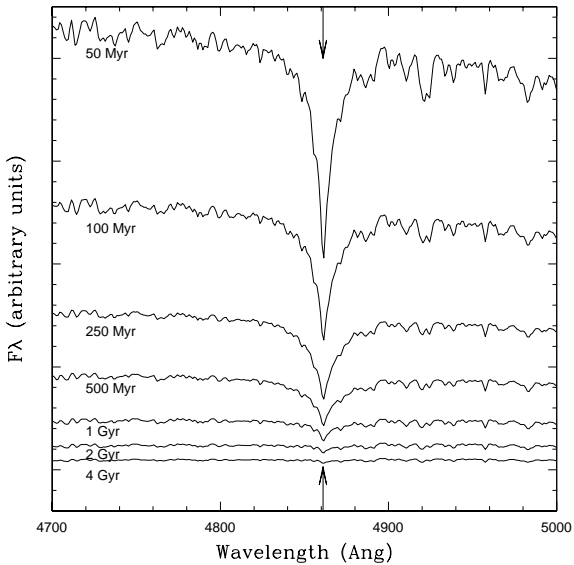}
\caption{
Simulated spectra showing the variation in the strength of the
H$\beta$ absorption feature for solar metallicity simple stellar populations, showing the extreme depth of the feature in young and intermediate-age stellar populations and hence the sensitivity of H$\beta$ spectroscopy to small fractional admixtures of such populations.}
\label{fig:SSP_spec}
\end{figure}

In order to derive useful properties, principally ages, from our spectra, we performed comparisons with theoretical spectra taken from the BaSTI stellar population synthesis data base \citep{piet04,perc09}. In all cases, the procedures adopted were identical for the observed and theoretical spectra, and involved direct integration of spectral flux.  EW values were measured using the LECTOR programme of A. Vazdekis\footnote{See http://www.iac.es/galeria/vazdekis/index.html} in bandpasses defined by \citet{trag98}.  Since the analysis was principally aimed at constraining ages of stellar populations, the most important index used was H$\beta$.  The strong dependence of this feature on stellar population age is shown in Fig.~\ref{fig:SSP_spec}, which clearly illustrates the monotonic decline in H$\beta$ strength with increasing age of a simple (single-age and single metallicity) stellar population (SSP), over the age range 50~Myr -- 4~Gyr.  
However, since H$\beta$ strength shows some dependence on metallicity, the Mgb index was also determined for all spectra to break a possible age-metallicity degeneracy when interpreting index strengths.

\subsection{Comparison with simple stellar population models}

\begin{figure}
\includegraphics[width=80mm,angle=0]{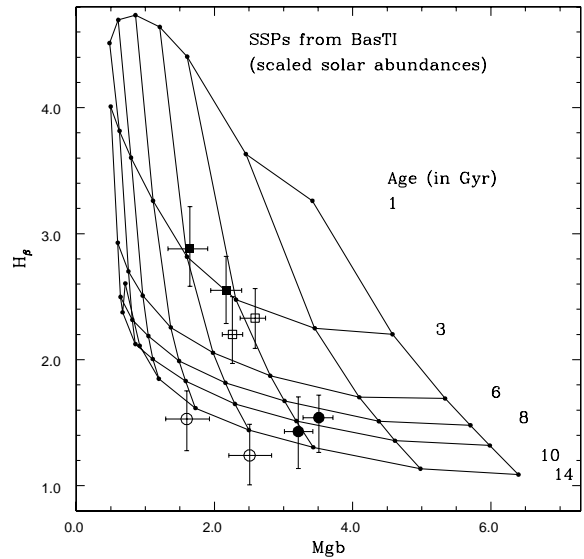}
\caption{
A grid of indices from BaSTI stellar population models, showing the strong 
correlation of H$\beta$ EW with age of a simple
stellar population.
The grid points and overlaid data points correspond to equivalent
widths of the appropriate features using the wavelength ranges defined 
in Trager et al. (1998). 
The data points are determined from spectra taken
in the `desert'
regions on opposite sides of the nucleus of the 4 barred galaxies
(solid square, NGC~2543; open square, NGC~2712; open circle, NGC~3185; solid circle, NGC~3351) and the H$\beta$ index strengths have been corrected for the presence of diffuse line emission.  SSP ages are indicated to the right of the figure; the metallicities used in calculating the grid points are [Fe/H] $=$ --1.27, --0.96, --0.66, --0.31, +0.06 and +0.4, from left to right.
}
\label{fig:SFD_sspgrid}
\end{figure}

For the initial comparison, we used SSP models to provide a useful reference point, while acknowledging that these poorly represent the extended and potentially complex SF histories of bright spiral galaxies.  The models used assumed scaled-solar element abundance ratios with total heavy element abundances ranging from well below to just above solar, and covered ages from 50~Myr to 14~Gyr.  The predicted H$\beta$ and Mgb index strengths for these SSPs are represented by the grid in Fig.~\ref{fig:SFD_sspgrid}, which shows the restricted age range of 1 -- 14~Gyr.

The observed spectra yielded H$\beta$ EWs covering the range 0.974 -- 2.334~\AA\, with the full set of values being listed in column 2 of Table~\ref{tab:hb_indices}. Such EWs are only compatible with SSPs of age$>$3~Gyr, regardless of metallicity, as is apparent from the model grid in Fig.~\ref{fig:SFD_sspgrid}. The lowest values of H$\beta$ EW are formally inconsistent with any model predictions, even for the oldest SSPs. Such low values motivate the study of contamination by emission lines, which can potentialy `fill in' the H$\beta$ absorption feature, leading to significant over-estimates of population ages.

\subsection{Correction for line emission within absorption features}

In JP15, we presented an analysis of these 4 galaxies which showed evidence for low-level diffuse emission lines, including H$\alpha$, throughout their SFD regions. Thus there must also be associated H$\beta$ emission, and the filling-in problem mentioned above must be occurring at some level.  However, the relative weakness of H$\beta$ emission compared with H$\alpha$, coupled with the strength of H$\beta$ absorption and the relatively poor spectral resolution of our WHT spectroscopy mean that no H$\beta$ emission is directly apparent in the reduced spectra of the SFD regions. In addition, the restricted spectral range covered by our data and the lack of contiguous coverage between the H$\alpha$ and H$\beta$ spectral regions make it difficult to fit a model to the continuum spectrum and hence directly decouple continuum and emission-line contributions, as is frequently done elsewhere.  Instead, we adopt an iterative procedure of corrections to the measured H$\alpha$ and H$\beta$ emission and absorption feature strengths to produce a self-consistent solution, as described below.  Iteration is needed as underlying absorption affects the measured H$\alpha$ line flux, while the H$\beta$ emission line flux affects the H$\beta$ absorption strength.

This iterative correction is implemented as follows.  First the H$\beta$ absorption EW is determined by direct integration on the observed spectrum in the bandpass defined by \citet{trag98} as described above.  The Mgb EW is derived in the same way, and the resulting index values are used to identify the closest matching SSP model spectrum.  This model spectrum is then used to derive an estimate of the H$\alpha$ absorption EW, which for practical purposes we define in a narrow 8~\AA\ passband.  This EW is then combined with the measured continuum flux density from the observed spectrum to derive a flux correction for the H$\alpha$ emission line, which was originally measured in a passband of width 8~\AA\ (JP15).  The corrected H$\alpha$ emission line flux is then used to predict the strength of the H$\beta$ emission line, assuming the widely-adopted Case B H$\alpha$/H$\beta$ ratio of 2.85 \citep[e.g.][]{broc71,humm87}, although this assumption is examined further below. The H$\beta$ absorption EW is then corrected for the effect of this predicted emission line, and the new EW used as the basis for the next stage of iteration. In practice, the process converged after 2 or 3 iterations.  The total sizes of the resulting corrections to the measured H$\beta$ EW
 values were between 0.18 and 0.62~\AA, which increased the strength of the H$\beta$ feature by between 13 and 32 per cent.  These corrected H$\beta$ indices are used in the discussion in the remainder of this paper, and are plotted in Fig.~\ref{fig:SFD_sspgrid} and subsequent figures. The measured H$\beta$ absorption indices, after emission-line correction, are listed in column 3 of Table~\ref{tab:hb_indices}.

\begin{table*}
 \begin{minipage}{140mm}
  \caption{Measured H$\beta$ absorption line indices, before and after emission-line correction.}
  \begin{tabular}{lcc}
  \hline
Galaxy and region &  H$\beta$, no corr.   &  H$\beta$, corrected  \\
                  &    \AA                &    \AA                \\
   \hline            
NGC~2543, SFD1    &  2.334    &  2.876 \\
NGC~2543, SFD2    &  1.921    &  2.544 \\
NGC~2712, SFD1    &  1.778    &  2.195 \\
NGC~2712, SFD2    &  1.812    &  2.326 \\
NGC~3185, SFD1    &  0.974    &  1.237 \\
NGC~3185, SFD2    &  1.233    &  1.527 \\
NGC~3351, SFD1    &  1.362    &  1.541 \\
NGC~3351, SFD2    &  1.244    &  1.429 \\
\hline
\end{tabular}
\label{tab:hb_indices}
\end{minipage} 
\end{table*}

The intrinsic H$\alpha$/H$\beta$ emission line ratio of 2.85 that was adopted above is used widely in the literature, but generally for regions ionized by hot stars. \citet{ho93}, following \citet{ferl83}, prefer a higher ratio of 3.1 for Low Ionization Nuclear Emission-line Region (LINER) type sources, noting that H$\alpha$ emission strength is boosted in regions dominated by collisional excitation.  Given that the SFD regions in these barred galaxies show LINER-like [NII]/H$\alpha$ ratios (JP15), these higher H$\alpha$/H$\beta$ ratios may be more appropriate here.  Adopting a ratio of 3.1 reduces the emission-line correction by about 10 per cent, resulting in corrected EW values that are lower by between 0.014 and 0.050~\AA. Extinction will also act to increase the effective H$\alpha$/H$\beta$ ratio, thus weakening this correction and decreasing the corrected absorption EW.  Extinction in these SFD regions is largely unconstrained, but we note that they are visually smooth and show no evidence of patchy dust obscuration, and so we apply no extinction correction.

Given the significant uncertainties in this emission-line correction process, we assume that it contributes an error equal to 50 per cent of the size of the final correction.  This error is much larger than, for example, the changes that would result from adopting any of the H$\alpha$/H$\beta$ intrinsic emission line ratios resulting from different stellar and non-stellar processes investigated by \citet{ferl83} and listed in their Table~3.

The absorption-corrected H$\beta$ indices, combined with Mgb metallicity-sensitive indices, are plotted in Fig.~\ref{fig:SFD_sspgrid}.  This shows that the index strengths are matched by SSPs with ages ranging between 3 and 14~Gyr. Fig.~\ref{fig:SFD_sspgrid} also indicates that the stellar populations appear to have moderately sub-solar metallicities, with most of the SFD points lying between [Fe/H] of +0.06 and --0.66.  This is slightly surprising for regions towards the centres of intrinsically luminous spiral galaxies, where metallicities similar to solar or somewhat higher might be expected. Given the shape of the SSP age-metallicity grid, shifting the points to somewhat higher metallicities would result in somewhat younger mean ages, but it is still clear that most of the measured points would indicate old stellar populations in these SFD regions, with equivalent SSP ages of several billion years.

\subsection{Extended SF histories with a defined truncation epoch}

\begin{figure}
\includegraphics[width=80mm,angle=0]{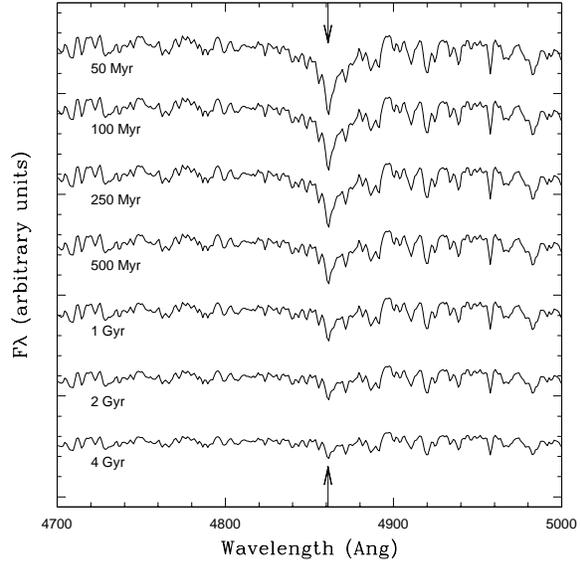}
\caption{
Simulated spectra showing the variation in the strength of the
H$\beta$ absorption feature as a function of the time since SF
activity was truncated, for solar metallicity populations.}
\label{fig:trunc_spec}
\end{figure}

One limitation of the above analysis was the comparison with SSP models, which will poorly represent the properties of any components of a real galaxy, where SF histories are likely to be temporally extended and complex.  In this section, we attempt to include, at least to first order, the effects of a more realistic SF history.  We do this by 
calculating index strengths from stellar populations with 
SF rates that are constant in time over an
interval starting 13~Gyr ago, and ending at a truncation epoch which
we vary from 4~Gyr to 50~Myr ago.  While such a model is clearly a simplification of the true history, this assumption gives a one-parameter set of models which is better suited to the specific question we address here than are, for example, the mass- and luminosity-weighted effective ages of stellar populations resulting from the more sophisticated full spectrum fitting of \citet{sanc11}. 

\begin{figure}
\includegraphics[width=80mm,angle=0]{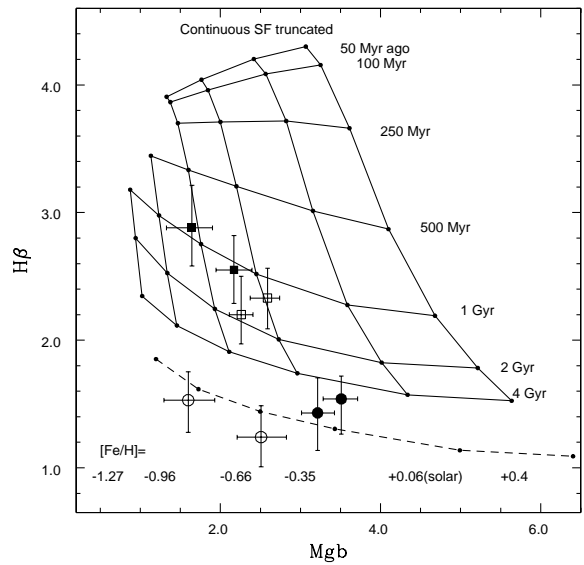}
\caption{
A grid of indices from BaSTI stellar population models, showing the strong 
correlation of H$\beta$ EW with time since truncation
of SF activity.
The grid points and overlaid data points correspond to equivalent
widths of the appropriate features using the same wavelength ranges,
defined in Trager et al. (1998). 
The data points are determined from spectra taken
in the SFD
regions on opposite sides of the nucleus of 4 barred galaxies
(Solid square, NGC~2543; open square, NGC~2712; open circle, NGC~3185; solid circle, NGC~3351), and the H$\beta$ index strengths have been corrected for the presence of diffuse line emission.
The dashed line at the bottom of the plot corresponds to SSPs of age 14~Gyr.
}
\label{fig:SFD_emcorpluserrors}
\end{figure}

Figure~\ref{fig:trunc_spec}
shows examples of the
modeled spectra, using the BaSTI models and spectral library \citep{perc09}.  The figure shows the decreasing depth of the
H$\beta$ feature as the truncation epoch is pushed back from 50~Myr
(top spectrum) to 4~Gyr (bottom).  This is the same trend in H$\beta$ strength with age as is shown in the SSP models (Fig.~\ref{fig:SSP_spec}), but with EW values overall being lower due to the effects of the oldest stars which are present in all models.
The grid of the H$\beta$ index vs
the Mgb index shown in Fig.~\ref{fig:SFD_emcorpluserrors}
quantifies the
variation in H$\beta$ with the epoch of last SF, showing
particularly strong variation over the range 0.1 - 2~Gyr.  This is
well matched to the requirements for discriminating between the
theoretical possibilities outlined in the introduction; if bars typically persist
for only a few rotations, the observations should populate the upper
part of the grid shown in Fig.~\ref{fig:SFD_emcorpluserrors}, whereas a continuation of
the current SF pattern for several Gyr should lead to an old
population in the SFD regions, populating the bottom of the diagram.  

These observations have yielded age estimates for the two SFD regions in each
of the four barred
galaxies, shown in Fig. 
\ref{fig:SFD_emcorpluserrors}.  All eight regions indicate that SF has
been curtailed for at least the last 1~Gyr, and hence that bars are
likely to be long-lived features.  However, significant differences are evident in the ages revealed by our spectroscopic analysis, with the SFD regions in two galaxies (NGC~3185 and NGC~3351) having very old populations, while NGC~2543 and NGC~2712 have evidence for more recent truncation of SF, within the last $\sim$2~Gyr. We discuss the consequences of these findings in the next section.

\section{Discussion}
\label{sec:discussion}

When discussing the implications of these implied SF histories, it is important to stress that we are studying the radial range which lies in the heart of the star-forming disk in unbarred galaxies, as illustrated by Fig.~1 of JP15. This is a key point for this analysis, and has not been addressed in previous studies of stellar populations and SF histories of barred galaxies. To make this point more quantitative, we have estimated the 
fraction of total SF that might be expected in an unbarred galaxy in the radial range covered by the extracted SFD slit spectra for each of the 4 galaxies studied. This was done using averaged H$\alpha$ profiles for  unbarred galaxies of the same Hubble $T$-type (Sb or Sa), taken from \citet{jame09}. The resulting fractions are: for NGC~2543, 33 per cent; for NGC~2712, 26 per cent; for NGC~3185, 37 per cent; and for NGC~3351, 15 per cent, with the smaller value for the last galaxy being due to its shorter bar.  Thus it is clearly not the case that we are simply determining the age of the old stellar population expected in bulge regions of galaxies; these observations cover a radial range that in unbarred but otherwise similar galaxies is characterised by disk regions with very vigorous SF activity.

The first finding from our spectroscopic analysis is that the stellar populations in SFD regions can be very old.  This is in agreement with the findings of \citet{sanc11} for four different barred galaxies, although their analysis methods were quite different.  Old stellar populations have also been found {\em within} bars by some studies \citep{pere07, pere09, delo12, delo13}, in general agreement with our results, but note that in this case it is not clear how to relate these ages to the bar age (an old population could have been assembled into a bar morphology very recently).

However, one of our galaxies is consistent with SF having been truncated only 1~Gyr ago, and another between 1 and 2 Gyr ago.  Thus we find a significant range in stellar population properties, even within a small sample of four early-type barred galaxies.  One advantage of the analysis method adopted is that we get two measurements of the SFD stellar populations within each galaxy, on opposite sides of the nucleus.  Given that the reduction of each region is completely independent of all the others (so that, for example, all of the parameters defining the emission-line correction are locally-derived and not averaged across both galaxy regions), it is reassuring to note that the line indices measured are entirely consistent for all of the pairs of spectra within individual galaxies. Thus, we do not find any indications of variation in stellar population properties within any of these four galaxies.  That is not true for the range of index strengths across the full sample of four galaxies, which is much larger than would be expected from measurement errors alone, indicating real differences in population ages or truncation epochs between these galaxies.

The galaxy sample studied here is too small for any formal statistical analysis, but it is instructive to look for correlations between SFD population ages and other galaxy properties.  The youngest population is found for the SFD of NGC~2543 (the two points lying on the 1~Gyr truncation age in Fig.~\ref{fig:SFD_emcorpluserrors}); this is classified as an SBb, with quite vigorous SF in the disk outside the SFD region.  The second youngest is found in NGC~2712, another SBb with very strong disk SF, that is morphologically very similar to NGC~2543.  There is then a significant jump to greater SFD age for NGC~3351, which is also classified as SBb.  The oldest population age is found for NGC~3185, the one galaxy in this sample classified as SBa, and the one that shows weakest (but still clearly detected) disk SF in its H$\alpha$ image (Fig.~\ref{fig:N3185}, lower frame).

Figures 1--4 show that these spectra sample regions that are significantly offset from the main bar axis.  However, there is some variation in this offset, with two galaxies observed with the slit perpendicular to the bar (NGC~2543 and NGC~3351) while NGC~2712 and NGC~3185 were observed at smaller angular offsets of 38$^{\circ}$ and 27$^{\circ}$ respectively.  This is potentially significant, given that the bar models of \citet{wozn07} and \citet{aume15} suggest that significant numbers of young stars may be trapped within the bar. However, there is no sign of such effects in the present sample, and indeed the youngest SFD population is found for NGC~2543, one of the cases where the slit was perpendicular to the bar and no contribution from the bar light can be expected.  Given the small size of the present sample, and the lack of spectra taken very close to the bar axis, we cannot make any strong statements about the predictions of \citet{wozn07} and \citet{aume15}, but this is an important question for future studies of this type.

While we confirm the result of JP15 that the SFD regions of the four galaxies studied have heavily suppressed SF rates, all four galaxies have strong emission-line peaks at their bar centres.  The observations presented in JP15 include the central regions of all four galaxies, although central properties were not discussed in that paper. The observed [NII]/H$\alpha$ central-region line ratios are consistent with expectations for SF in the nucleus of one galaxy (NGC~2543), in central rings in two others (NGC~2712 and NGC~3185), and in both nucleus and central ring for the remaining galaxy (NGC~3351). This confirms the general finding of other studies \citep[e.g.][]{pere00, coel11, elli11} that bars can efficiently induce SF in the central regions of their host galaxies, and we show here that such SF can co-exist with the SFD phenomenon.

Many studies have discussed the possibility of radial migration of stars.  This was first introduced as an effect that is driven by spiral arm patterns at their corotation resonances \citep{sell02,rosk08,loeb11}. Other studies have explored the possible effects of bar--spiral arm coupling on radial migration \citep{deba06, brun11, minc11, minc12, dima13}, although the evidence for strong bar-driven migration is not widely accepted \citep{rosk12}, and \citet{sanc14} argue against strong radial mixing in their study of metallicity gradients in barred and unbarred galaxies. If such radial mixing {\em does} occur, it would complicate the interpretation of the results presented here.  However, any mixing of the young stellar populations from either the central SF peak or the surrounding SF ring into the SFD region would only strengthen the general result we find here, of long-suppressed SF activity within the SFD region itself.

One obvious extension of the analysis presented here would be to look at the stellar populations in the disk outside the SFD, but between HII regions, to determine the characteristic age of the non-star-forming disk population.  This was attempted with the present long-slit data, but did not prove feasible, as any line-free regions were very small and of such low surface brightness that it was impossible to fit any absorption line features in any such disk regions.  This demonstrates the unique nature of the SFD regions as extensive, relatively high surface brightness areas that are free of ongoing star formation.  The analysis of outer disk regions might be possible using integral field units, through the co-addition of spectra covering large areas with the removal of all those showing line emission.

As noted in JP15, the analysis presented here only applies to a particular type of bar, that is
characteristic of early-type spirals and is not found in late types.
Future areas of study include extending this type of analysis to such 
later type strongly barred galaxies, to determine whether the bimodality
in bar types noted by, e.g., \citet{nair10} and \citet{hako14} is reflected 
in the associated stellar populations. It would also be useful to apply the 
methods of the present study to the stellar population found within the 
body of the bar itself.  Future studies might also be able to constrain radial 
migration effects by looking for age gradients in the outer parts of the SFD, as the star-forming ring is approached.

\section{Conclusions}
\label{sec:conc_fut_work}

The principal conclusions of this study are as follows:

\begin{itemize}
\item We confirm the SF pattern found in our earlier study, for four strongly-barred galaxies, with a central peak and an outer ring both showing strong activity, and an intermediate `desert' where SF is strongly suppressed.
\item It is important to correct appropriately for low-level diffuse line emission, identified in all four of these galaxies; failure to do this results in significantly overestimated population ages.
\item We derive two independent estimates of SFD ages per galaxy, which are internally consistent for each of the four galaxies.
\item Overall we find old stellar populations in SFD regions, indicate that SF was truncated at least several dynamical times ago.
\item However, there is a large range in ages or truncation epochs found across the sample of four galaxies, not consistent with expectations from measurement errors alone.
\item Further work is required to extend this analysis to statistically significant samples of galaxies, covering a wider range of galaxy types, and to constrain complicating effects such as radial migration of stars.
\end{itemize}

\section*{Acknowledgments}

The William Herschel and Isaac Newton Telescopes are operated on the island of La Palma by the Isaac Newton Group in the Spanish Observatorio del Roque de los Muchachos of the Instituto de Astrof\'isica de Canarias.
This research has made use of the NASA/IPAC Extragalactic Database (NED) which is operated by the Jet Propulsion Laboratory, California Institute of Technology, under contract with the National Aeronautics and Space Administration. 
SMP acknowledges the award of Visiting Researcher status by Liverpool John Moores University.


\begin{thebibliography}{}


\bibitem[\protect\citeauthoryear{Aumer \& Sch{\"o}nrich}{2015}]{aume15} Aumer M., Sch{\"o}nrich R., 2015, MNRAS, 454, 3166 

\bibitem[\protect\citeauthoryear{Berentzen et al.}{2007}]{bere07} 
Berentzen I. et al., 2007, ApJ, 666, 189

\bibitem[\protect\citeauthoryear{Bournaud \& Combes}{2002}]{bour02} 
Bournaud F., Combes, F. 2002, A\&A, 392, 83

\bibitem[\protect\citeauthoryear{Bournaud, Combes \& Semelin}{2005}]{bour05} 
Bournaud F., Combes, F., Semelin, B., 2005, MNRAS, 364, L18

\bibitem[\protect\citeauthoryear{Brocklehurst}{1971}]{broc71} 
Brocklehurst M., 1971, MNRAS, 153, 471 

\bibitem[\protect\citeauthoryear{Brunetti, Chiappini, \& Pfenniger}{2011}]{brun11} Brunetti M., Chiappini C., Pfenniger D., 2011, A\&A, 534, A75 

\bibitem[\protect\citeauthoryear{Coelho 
\& Gadotti}{2011}]{coel11} Coelho P., Gadotti D.~A., 2011, ApJ, 743, L13 

\bibitem[\protect\citeauthoryear{Combes et 
al.}{1990}]{comb90} Combes F., Debbasch F., Friedli D., Pfenniger D., 1990, A\&A, 233, 82 

\bibitem[\protect\citeauthoryear{Curir, Mazzei, \& Murante}{2008}]{curi08} Curir A., Mazzei P., Murante G., 2008, A\&A, 481, 651 

\bibitem[\protect\citeauthoryear{de Lorenzo-C{\'a}ceres, 
Falc{\'o}n-Barroso, \& Vazdekis}{2013}]{delo13} 
de Lorenzo-C{\'a}ceres A., Falc{\'o}n-Barroso J., Vazdekis A., 2013, MNRAS, 431, 2397 

\bibitem[\protect\citeauthoryear{de Lorenzo-C{\'a}ceres et al.}{2012}]{delo12} de Lorenzo-C{\'a}ceres A., Vazdekis A., Aguerri J.~A.~L., Corsini E.~M., Debattista V.~P., 2012, MNRAS, 420, 1092

\bibitem[\protect\citeauthoryear{Debattista et al.}{2006}]{deba06} Debattista V.~P., Mayer L., Carollo C.~M., Moore B., Wadsley J., Quinn T., 2006, ApJ, 645, 209 

\bibitem[\protect\citeauthoryear{Di Matteo et al.}{2013}]{dima13} Di Matteo P., Haywood M., Combes F., Semelin B., Snaith O.~N., 2013, A\&A, 553, A102 

\bibitem[\protect\citeauthoryear{Ellison et al.}{2011}]{elli11} Ellison S.~L., Nair P., Patton D.~R., Scudder J.~M., Mendel J.~T., Simard L., 2011, MNRAS, 416, 2182 


\bibitem[\protect\citeauthoryear{Ferland \& Netzer}{1983}]{ferl83} Ferland G.~J., Netzer H., 1983, ApJ, 264, 105

\bibitem[\protect\citeauthoryear{Gil de Paz et al.}{2007}]{gild07} Gil de Paz A., et al., 2007, ApJS, 173, 185

 
\bibitem[\protect\citeauthoryear{Hakobyan et al.}{2014}]{hako14} Hakobyan A.~A., et al., 2014, MNRAS, 444, 2428 

\bibitem[\protect\citeauthoryear{Hakobyan et al.}{2015}]{hako15} Hakobyan A.~A., et al., 2015, arXiv, arXiv:1511.08896 

\bibitem[\protect\citeauthoryear{Ho, Filippenko, \& Sargent}{1993}]{ho93} Ho L.~C., Filippenko A.~V., Sargent W.~L.~W., 1993, ApJ, 417, 63 

\bibitem[\protect\citeauthoryear{Hozumi}{2012}]{hozu12} Hozumi S., 2012, PASJ, 64, 5 

\bibitem[\protect\citeauthoryear{Hozumi \& Hernquist}{2005}]{hozu05} Hozumi S., Hernquist L., 2005, PASJ, 57, 719 

\bibitem[\protect\citeauthoryear{Hummer \& Storey}{1987}]{humm87} Hummer D.~G., Storey P.~J., 1987, MNRAS, 224, 801


 
\bibitem[\protect\citeauthoryear{James, Bretherton, \& Knapen}{2009}]{jame09} James P.~A., Bretherton C.~F., Knapen J.~H., 2009, A\&A, 501, 207 

\bibitem[\protect\citeauthoryear{James 
\& Percival}{2015}]{jame15} James P.~A., Percival S.~M., 2015, MNRAS, 450, 3503 (JP15) 

 
\bibitem[\protect\citeauthoryear{Knapen et al.}{2000}]{knap00} 
Knapen J.~H., Shlosman I. \& Peletier R., 2000, ApJ, 529, 93
 
\bibitem[\protect\citeauthoryear{Kormendy \& Kennicutt}{2004}]{korm04} 
Kormendy J., Kennicutt R.~C., 2004, ARAA, 42, 603

\bibitem[\protect\citeauthoryear{Lang, Holley-Bockelmann, 
\& Sinha}{2014}]{lang14} Lang M., Holley-Bockelmann K., Sinha M., 2014, ApJ, 790, L33

\bibitem[\protect\citeauthoryear{Loebman et al.}{2011}]{loeb11} Loebman S.~R., Ro{\v s}kar R., Debattista V.~P., Ivezi{\'c} {\v Z}., Quinn T.~R., Wadsley J., 2011, ApJ, 737, 8 
 
\bibitem[\protect\citeauthoryear{Marinova \& Jogee}{2007}]{mari07} 
Marinova I., Jogee S., 2007, ApJ, 659, 1176 
 
\bibitem[\protect\citeauthoryear{Minchev et al.}{2011}]{minc11} Minchev I., Famaey B., Combes F., Di Matteo P., Mouhcine M., Wozniak H., 2011, A\&A, 527, A147 


\bibitem[\protect\citeauthoryear{Minchev et al.}{2012}]{minc12} Minchev I., Famaey B., Quillen A.~C., Di Matteo P., Combes F., Vlaji{\'c} M., Erwin P., Bland-Hawthorn J., 2012, A\&A, 548, A126 

\bibitem[\protect\citeauthoryear{Nair \& Abraham}{2010}]{nair10} Nair P.~B., Abraham R.~G., 2010, ApJ, 714, L260

\bibitem[\protect\citeauthoryear{Percival et al.}{2009}]{perc09} 
Percival S.~M., Salaris M., Cassisi S. \& Pietrinferni A., 2009, ApJ, 690, 427 
 

\bibitem[\protect\citeauthoryear{P{\'e}rez, S{\'a}nchez-Bl{\'a}zquez, 
\& Zurita}{2007}]{pere07} P{\'e}rez I., S{\'a}nchez-Bl{\'a}zquez P., Zurita A., 2007, A\&A, 465, L9 

\bibitem[\protect\citeauthoryear{P\'erez, S{\'a}nchez-Bl{\'a}zquez, 
\& Zurita}{2009}]{pere09} P\'erez I., S\'anchez-Bl\'azquez P. \& Zurita, A., 2009, A\&A, 495, 775

\bibitem[\protect\citeauthoryear{P{\'e}rez \& S{\'a}nchez-Bl{\'a}zquez}{2011}]{pere11} P{\'e}rez I., S{\'a}nchez-Bl{\'a}zquez P., 2011, A\&A, 529, A64

\bibitem[\protect\citeauthoryear{P\'erez-R\'amirez et al.}{2000}]{pere00} P\'erez-R\'amirez D., Knapen J.~H., Peletier R.~F., Laine S., Doyon R., Nadeau D., 2000, MNRAS, 317, 234 

\bibitem[\protect\citeauthoryear{Pietrinferni et al.}{2004}]{piet04} Pietrinferni A., Cassisi S., Salaris M., Castelli F., 2004, ApJ, 612, 168 

\bibitem[\protect\citeauthoryear{Reynaud \& Downes}{1998}]{reyn98} Reynaud D., Downes D., 1998, A\&A, 337, 671 

\bibitem[\protect\citeauthoryear{Ro{\v s}kar et al.}{2008}]{rosk08} Ro{\v s}kar R., Debattista V.~P., Quinn T.~R., Stinson G.~S., Wadsley J., 2008, ApJ, 684, L79 

\bibitem[\protect\citeauthoryear{Ro{\v s}kar et al.}{2012}]{rosk12} Ro{\v s}kar R., Debattista V.~P., Quinn T.~R., Wadsley J., 2012, MNRAS, 426, 2089

\bibitem[\protect\citeauthoryear{S{\'a}nchez et al.}{2014}]{sanc14} S{\'a}nchez S.~F., et al., 2014, A\&A, 563, A49 

\bibitem[\protect\citeauthoryear{S{\'a}nchez-Bl{\'a}zquez et al.}{2011}]{sanc11} S{\'a}nchez-Bl{\'a}zquez P., Ocvirk P., Gibson B.~K., P{\'e}rez I., Peletier R.~F., 2011, MNRAS, 415, 709 




\bibitem[\protect\citeauthoryear{Sellwood \& Binney}{2002}]{sell02} Sellwood J.~A., Binney J.~J., 2002, MNRAS, 336, 785 

\bibitem[\protect\citeauthoryear{Shen \& Sellwood}{2004}]{shen04} Shen J., Sellwood J.~A., 2004, ApJ, 604, 614 

\bibitem[\protect\citeauthoryear{Singh et al.}{2013}]{sing13} Singh R., et al., 2013, A\&A, 558, AA43

\bibitem[\protect\citeauthoryear{Trager et al.}{1998}]{trag98} Trager S.~C., Worthey G., Faber S.~M., Burstein D., Gonz{\'a}lez J.~J., 1998, ApJS, 116, 1

\bibitem[\protect\citeauthoryear{Wozniak}{2007}]{wozn07} Wozniak H., 2007, A\&A, 465, L1 


\end{thebibliography}

\label{lastpage}

\end{document}